\documentclass{elsart}

\usepackage{amssymb}
\usepackage{amsmath}

\usepackage{graphicx}

\makeatletter
\def\pubpagerange#1{\gdef\@pubpagerange{#1}}
\def\@jou@vol@pag{\@journal\ \@volume\ (\the\@pubyear)\ \@pubpagerange}
\makeatother
  
\journal{Physica C}
\volume{468}
\pubyear{2008}
\pubpagerange{2403-2407}

\begin{document}

\begin{frontmatter}



\title{Anisotropic superconductivity in graphite intercalation compound YbC$_{6}$}


\author{N.F. Kawai\corauthref{cor}},
\corauth[cor]{Corresponding author. Tel.: +81 3 5841 8364; fax: +81 3 5841 8364.}
\ead{kawai@kelvin.phys.s.u-tokyo.ac.jp}
\author{Hiroshi Fukuyama}

\address{Department of Physics, Graduate School of Science, The University of Tokyo, 
7-3-1 Hongo, Bunkyo-ku, Tokyo 113-0033, Japan}

\begin{abstract}
We report anisotropy of the upper critical field ($B_{c2}$) of 
an intercalated graphite superconductor YbC$_{6}$ ($T_{c}$ = 6.5 K) 
determined from angular dependent magnetoresistance measurements. 
Though the perpendicular coherence length is much longer than 
interlayer spacing, measured angular dependences of $B_{c2}$ are 
well fitted by the Lawrence-Doniach model or the Tinkham model, 
which are known to be applicable to quasi two-dimensional 
materials or thin films, rather than the effective mass model. 
This observation is similar to the measurements for the other 
intercalated graphite superconductor, CaC$_{6}$, by Jobiliong et al. 
[E. Jobiliong, H.D. Zhou, J.A. Janik, Y.-J. Jo, L. Balicas, 
J.S. Brooks, C.R. Wiebe, Phys. Rev. B 76 (2007) 31 052511]. 
A possible explanation for the unexpected applicability of 
these models is that our YbC$_{6}$ samples are synthesized 
as thin flakes in the host graphite.
\end{abstract}

\end{frontmatter}





Graphite intercalation compounds (GICs) have been intensively 
studied for many years. This is because the electronic and 
magnetic properties can be controlled to a large extent by changing 
intercalant species and their arrangements such as the stage number and 
the intralayer structure \cite{r1}. Superconductivity is one of the most 
interesting properties among them. The superconducting transition 
temperature ($T_{c}$) of GICs had been long limited below 3 K under ambient 
pressure \cite{r1}.  The intercalants here are usually alkali 
metals. Recently, it was discovered that CaC$_{6}$ and YbC$_{6}$ have much 
higher $T_{c}$ of 11.5 and 6.5 K, respectively \cite{r2}. 
Measured properties of these new {\it high-}$T_{c}$ GICs are well described 
with the framework of BCS theory \cite{r3, r4, r5, r6, 
r7}. Although the phonon mechanism is supposed to be relevant 
for their Cooper pairing \cite{r8}, detailed microscopic mechanisms 
for the attractive interactions have not been fully understood yet.

Since GICs are structurally anisotropic, the superconducting 
properties also show anisotropic features. For instance, the upper 
critical field ($B_{c2}$) of KHgC$_{8}$ shows angular dependences 
which are well fitted by the effective mass (EM) model \cite{r9}. 
On the other hand, the anisotropy of $B_{c2}$ in CaC$_{6}$ is 
accounted for by the Lawrence-Doniach (LD) model or the Tinkham model 
\cite{r10}, though these models are applicable mainly to quasi 
two-dimensional (2D) materials or thin films. Here we report measurements 
of angular dependent magnetoresistance in YbC$_{6}$. The anisotropy of $B_{c2}$ 
deduced from our data is well described by the LD model or the 
Tinkham model. We infer that this result comes from a flake-shaped 
distribution of YbC$_{6}$ in the host graphite in our sample.

The YbC$_{6}$ samples were synthesized from the host material, a highly 
oriented pyrolytic graphite (HOPG) \cite{r11}, by the vapor transport 
method. A pellet of Yb and a piece of HOPG were vacuum-sealed in a 
glass ample, and heat-treated at 400 $^\circ$C for 2 days. After 
cooling down, the glass ample was opened inside a glove box with 
N$_{2}$ flow, because YbC$_{6}$ is extremely reactive to air. 
In the case of samples for transport measurements, Au lead wires 
were attached on HOPG with Ag paste before the heating. The typical 
sample size is about 5$\times$2$\times$0.3 mm$^{3}$.

The synthesized samples were first analyzed by X-ray diffraction 
(XRD) with CuK$\alpha$ radiation at room temperature. The sample surfaces 
for the XRD measurements were protected with adhesive tapes against 
reaction in air. Fig. 1 is a measured XRD spectrum for such a sample 
showing both YbC$_{6}$ (00l) peaks and host graphite ones. 
The interlayer spacing ($d'$) deduced from this spectrum is 0.456 nm 
which is almost the same as the value reported by Weller et al. \cite{r2}.

DC magnetization measurements in magnetic fields ($B$) parallel to the 
$ab$-plane of YbC$_{6}$ were carried out with a SQUID magnetometer \cite{r12}. 
The inset of Fig. 2 shows temperature dependences of magnetization ($M$) 
measured at $B$ = 1 mT by field cooling (FC, closed circles) and zero 
field cooling (ZFC, open circles). It shows $T_{c}$ = 6.5 K, and an 
estimated superconducting volume fraction is a few percent. 
Field dependences of $M$ measured at various fixed temperatures are 
shown in Fig. 2. In this plot, linear background signals, which are 
at most 6$\times$10$^{-3}$ emu at $B$ = 1 T, have already been subtracted 
from the raw data. The origin of this background signal is not known 
at present. At each temperature, the upper critical field in the parallel 
direction ($B_{c2\parallel}$) is determined as a field where magnetization 
returns to zero as indicated by the arrows in the figure.

Resistance measurements across the $ab$-plane were carried out using the standard 
four-probe method \cite{r13} down to $T$ = 2 K. A measurement current was 
fixed at 1 mA. Sample surfaces for the transport measurements were protected 
with epoxy resin against reaction in air. Fig. 3 shows the temperature 
dependence of resistance ($R$) in zero magnetic field. The $T_{c}$ value 
(= 6.47 K) here is defined as $R$($T_{c}$) =  $\frac{1}{2}$ $R$($T$ = 6.6 K) 
(see the inset of Fig. 3). This is consistent with the $T_{c}$ value determined 
from the magnetization measurements. The transition width corresponding to 
a resistance change from 0.1 to 0.9$R$($T$ = 6.6 K) is 0.1 K. 
Note that $R$ decreases to zero at $T$ $\le$ 6.4 K, implying that 
superconducting paths connect the two voltage leads in spite of 
the small volume fraction of YbC$_{6}$ in the host. 

Field dependences of resistance measured at various fixed temperatures 
are shown in Fig. 4. The field is applied parallel to the $ab$-plane. 
At each temperature, the resistance increases rather sharply above 
a certain field which corresponds to $B_{c2\parallel}$ as will be 
discussed later. With further increasing field, it approaches gradually 
to a universal curve ($R_{n}$($B$)) in the normal state whose apparent 
field dependence is governed by the host material. Note that the 
magnetoresistance of graphite is very large \cite{r14}.

$B_{c2\parallel}$ values determined in this way are plotted as a function 
of temperature by the closed circles in Fig. 5. $B_{c2\parallel}$ varies 
linearly with $T$ in the measured temperature range (2 K $\le$ $T$ $\le$ $T_{c}$). 
There are in good agreement with those determined from the magnetization 
measurements shown by the crosses in the figure. This consistency 
suggests appropriateness of the two independent determinations of $B_{c2\parallel}$. 
The open circles are fields where $R$($B$) = $\frac{1}{2}$ $R_{n}$($B$). 
This is not an appropriate definition for $B_{c2\parallel}$, since they are 
substantially (by $\sim$ 65 $\%$) larger than the $B_{c2\parallel}$ values 
determined from the magnetization measurements.

Next, we show results of resistance measurements in magnetic fields of 
various directions in Fig. 6. We define $\theta$ as an angle between 
the field direction and the $c$-axis of YbC$_{6}$. As $\theta$ 
decreases from 90$^\circ$ to 0$^\circ$, the field dependence of 
$R_{n}$($B$) becomes steeper reflecting the anisotropic magnetoresistance 
of host graphite \cite{r14}. From these data, temperature dependences of 
$B_{c2}$ in various field directions are obtained (Fig. 7). The $T$-linear 
dependence of $B_{c2}$ holds for any $\theta$. 
Anisotropy parameter ($\varepsilon \equiv B_{c2 \parallel}/B_{c2 \perp}$) is 
calculated to be 2.1 for $T$ = 2 K.
Angular dependences 
of $B_{c2}$ at various fixed temperatures are also obtained from the 
data in Fig. 6 and plotted in Fig. 8. At any temperature, $B_{c2}$ 
is anisotropic and a cusp is observed around $\theta$ = 90$^\circ$. 

We now compare the measured properties of YbC$_{6}$ 
with three different theoretical models for anisotropic 
superconductivity in layered materials or thin films. 
The first one is the EM model \cite{r15}, which introduces 
the anisotropy of the effective mass into the GL equations. 
This model deals superconductors as anisotropic continua, and is 
relevant when the coherence length perpendicular to layers 
is much longer than the interlayer spacing. In this model, 
the angular dependence of $B_{c2}$ is given by
\begin{equation}\label{eq1}
\left(\frac{B_{c2}(\theta) \sin \theta}{B_{c2 \parallel}}\right)^2 + \left(\frac{B_{c2}(\theta) \cos \theta}{B_{c2 \perp}}\right)^2 = 1,
\end{equation}
where $B_{c2\perp}$ is the upper critical field in perpendicular direction. 
Also, relations between $B_{c2}$ and the coherence length ($\xi$) are given by
\begin{alignat}{10}
&\quad B_{c2 \perp}(T) &&= \frac{\Phi_0}{2\pi \xi_{\parallel}(T)^2} &&= \frac{\Phi_0}{2\pi \xi_{\parallel}(0)^2}\left(1- \frac{T}{T_c} \right) &&\quad(T \approx T_c)\quad\label{eq2}\\
\!\!\!\!\!\!\!\mbox{and}&&&&&&&\nonumber\\
&\quad B_{c2 \parallel}(T) &&= \frac{\Phi_0}{2\pi \xi_{\parallel}(T) \xi_{\perp}(T)} &&= \frac{\Phi_0}{2\pi \xi_{\parallel}(0) \xi_{\perp}(0)}\left(1- \frac{T}{T_c} \right) &&\quad(T \approx T_c).\quad\label{eq3}
\end{alignat}
Here $\Phi_{0}$ = $h/2e$ is the flux quantum, and 
$\xi_{\perp}$ and $\xi_{\parallel}$ are coherence lengths 
perpendicular and parallel to layers, respectively. From Eq. (\ref{eq2}), 
Eq. (\ref{eq3}) and the data in Fig. 7, $\xi_{\perp}$ and $\xi_{\parallel}$ 
are estimated as 
\begin{alignat}{10}
&\xi_{\perp}(0) &&= 17 \ \mathrm{nm},\label{eq4}\\
&\xi_{\parallel}(0) &&= 37 \ \mathrm{nm}.\label{eq5}
\end{alignat}
This $\xi_{\perp}$(0) value is much larger than the interlayer spacing 
$d'$ = 0.456 nm, which does not exclude the applicability of the 
EM model. However, as shown in Fig. 8, fitting of the angular 
dependences of $B_{c2}$ to Eq. (\ref{eq1}) is poor (dashed lines). 
Particularly, this model does not represent the cusps around $\theta$ = 90$^\circ$.

There are two other models which may be applicable to our data. 
They are the LD model \cite{r16} and the Tinkham model \cite{r17}. 
The LD model describes anisotropic superconductors as stacked 2D 
superconducting layers connected by Josephson coupling. On the other 
hand, the Tinkham model is to describe superconducting thin films. 
In both models, the angular dependence of $B_{c2}$ is given by
\begin{equation}\label{eq6}
\Bigl|\frac{B_{c2}(\theta) \sin \theta}{B_{c2 \parallel}}\Bigr| + \left(\frac{B_{c2}(\theta) \cos \theta}{B_{c2 \perp}}\right)^2 = 1.
\end{equation}
These two models provide much better fittings for the measured 
angular dependences in Fig.8 (solid lines). This suggests the 
thin film nature of our sample. Since $\xi_{\perp}$ $\gg$ $d'$, 
it would be hard to apply the LD model to the present results. 
Instead, YbC$_{6}$ might be synthesized as thin flakes in the 
host acting as independent superconducting thin films. This is 
conceivable for the following reason. The host HOPG contains 
high-density stacking faults, and its effective thickness is 
estimated to be around 40 layers ($\sim$ 13 nm) \cite{r18}. 
So, it is likely that the thickness of synthesized YbC$_{6}$ 
is also restricted by this layer number.

Although the Tinkham model is originally developed for the case of 
$d < \xi_{\perp}$ \cite{r17}, it is known to be a fairly good 
approximation even for the case of $d \sim \xi_{\perp}$ where 
$\varepsilon \sim 2$ unless $d \gg \xi_{\perp}$ \cite{rev1_r1}. 
Therefore, it seems reasonable that the angular dependences of $B_{c2}$ 
obey Eq. (\ref{eq6}), while $\varepsilon$ (= 2.1) is not so large. 
The same angular dependences of $B_{c2}$ were also reported by 
Jobiliong et al. \cite{r10} for another {\it high}-$T_{c}$ 
GIC superconductor CaC$_{6}$ ($T_{c}$ = 11.5 K), 
where $\varepsilon = 3.7 \sim 4.6$. 

With the Tinkham model, other measured properties of our sample 
can be explained. First, a possible thickness variation in the 
superconducting paths would broaden the resistive transition in 
magnetic fields, since $B_{c2}$ increases with decreasing 
thickness in thin film superconductors \cite{r19}. Next, 
the fact that the $B_{c2}$ values measured in this work are 
larger than those by Weller et al. \cite{r2} for 
$\theta$ = 90$^\circ$ (see Fig. 7) may be also due to 
the thinner effective thickness of our sample. However, such a 
difference is not seen at $T$ $\ge$ 4 K for $\theta$ = 0$^\circ$, 
for which we don't have reasonable explanation at this moment.

There remain a few points which are not accounted for by the 
Tinkham model. For instance, the measured $B_{c2}$ at 
$\theta$ = 90$^\circ$ shows $T$-linear dependence rather 
than $\sqrt{T_{c}-T}$ dependence
\begin{equation}\label{eq7}
B_{c2}(T) = \sqrt{12}\frac{\Phi_0}{2\pi \xi d} \propto \sqrt{T_c - T}
\end{equation}
predicted by this model. Here, $d$ is film thickness. The same 
disagreement has been reported 
by Jobiliong et al. \cite{r10} 
even for a bulk GIC superconductor, i.e., CaC$_{6}$ 
synthesized from HOPG 
by the direct reaction method \cite{r10}. On the other hand, for YbC$_{6}$, 
the temperature dependence of $B_{c2}$ at $\theta$ = 90$^\circ$ 
measured by Weller et al. \cite{r2} is not linear with $T$ nor 
$\sqrt{T_{c}-T}$. In order to solve this puzzle, future synthesis 
of YbC$_{6}$ from single crystal graphite in well-defined form 
such as bulk is highly desirable.

In conclusion, we synthesized YbC$_{6}$, a superconducting GIC with 
$T_{c}$ = 6.5 K, by the vapor transport method. The measured angular 
dependences of $B_{c2}$ at various field directions are well fitted 
by the Lawrence-Doniach (LD) model or the Tinkham model. This indicates 
the thin film nature of sample, which might originate from the thin 
film nature of host HOPG with high-density stacking faults. This 
scenario including the currently unexplained temperature dependence 
of $B_{c2}$ will be tested by synthesizing sample with better defined 
form in future.

\section*{Acknowledgments}

We would like to thank C. Winkelmann and H. Kambara for their contributions 
in the early stage of this work. We would also like to thank S. Uchida 
and J. Shimoyama for their helpful suggestions in the magnetization and XRD measurements, 
and N. Akuzawa, Y. Iye and T. Matsui for valuable discussions. This work was 
financially supported by Grant-in-Aid for Scientific Research on 
Priority Areas (No. 17071002) and Global COE Program ``the Physical 
Sciences Frontier'' from MEXT, Japan.

\newpage

\begin{figure}[h]
\begin{center}
\includegraphics[height=80mm]{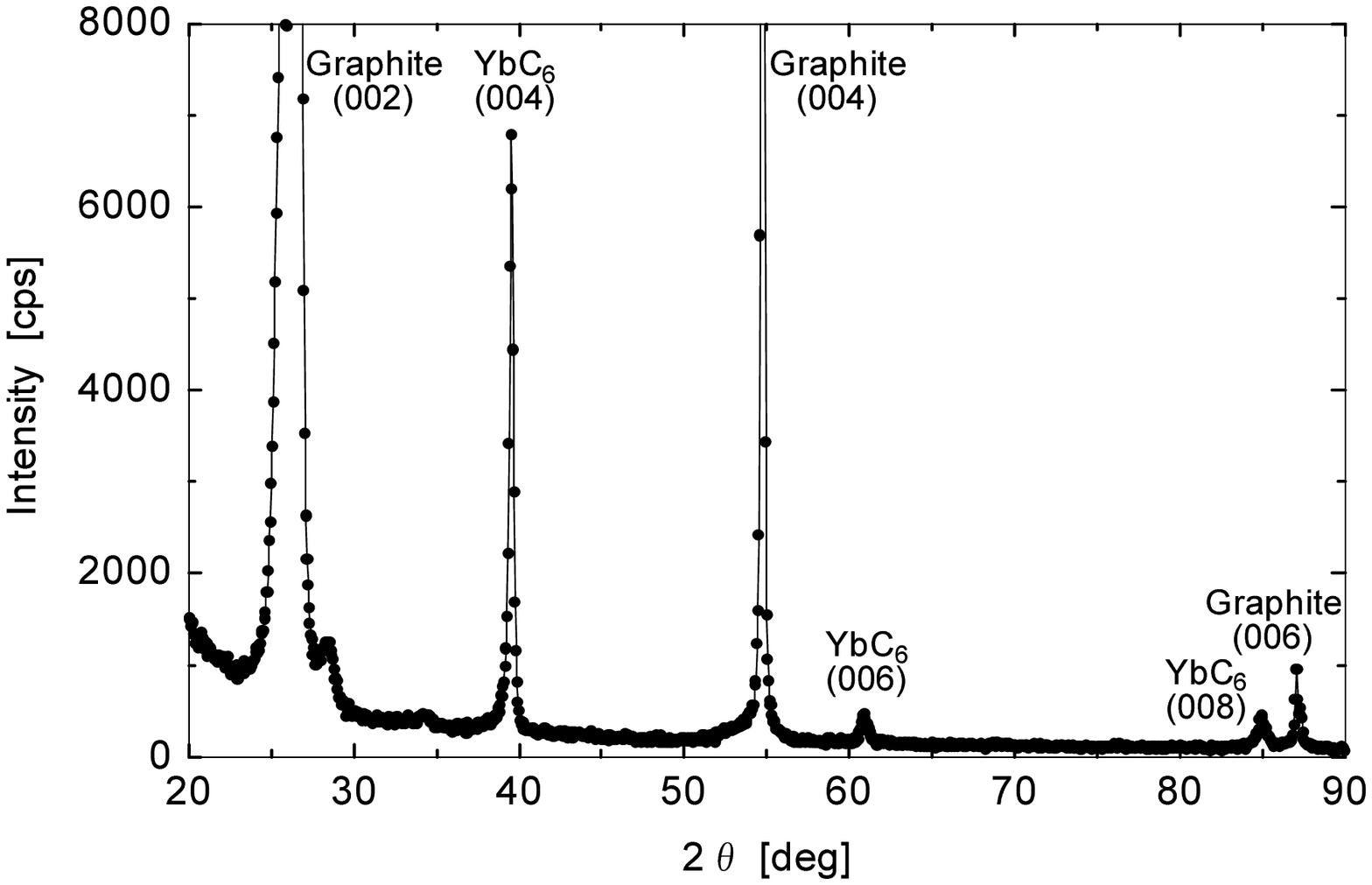}
\end{center}
\caption{The XRD spectrum of a synthesized YbC$_{6}$ sample measured with 
CuK$\alpha$ radiation at room temperature. Only (00l) peaks are 
observed because of the $c$-axis orientation of both host HOPG and YbC$_{6}$.}
\label{fig1}
\end{figure}

\newpage

\begin{figure}[h]
\begin{center}
\includegraphics[height=80mm]{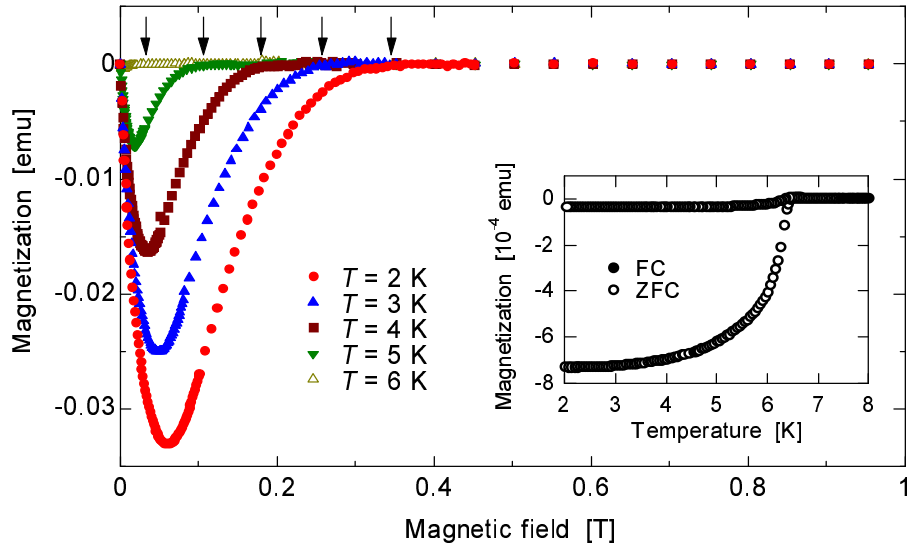}
\end{center}
\caption{The field dependence of magnetization in YbC$_{6}$. 
The field direction is parallel to the $ab$-plane. Linear background 
signals have already been subtracted from the raw data. 
$B_{c2\parallel}$ values are indicated by the arrows. Inset: 
the temperature dependence of magnetization measured at $B$ = 1 mT 
($B$ $\parallel$ $ab$-plane) with field cooling (FC) and zero field 
cooling (ZFC).}
\label{fig2}
\end{figure}

\newpage

\begin{figure}[h]
\begin{center}
\includegraphics[height=80mm]{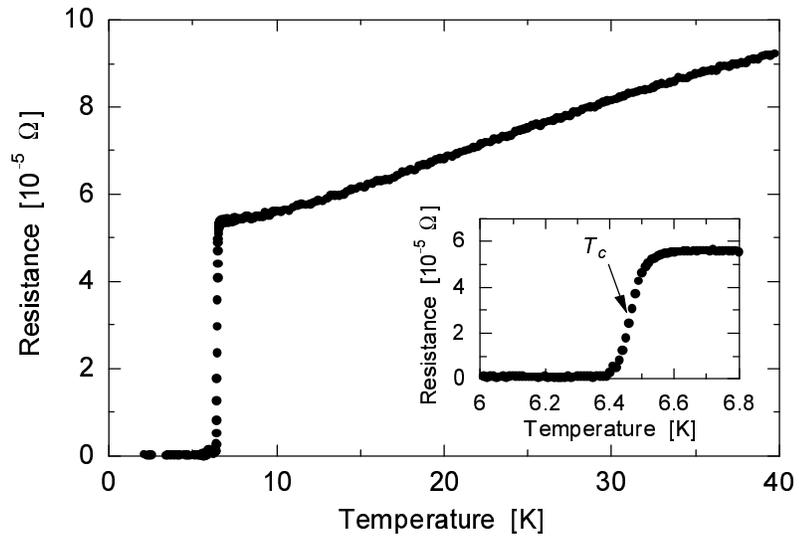}
\end{center}
\caption{The temperature dependence of resistance for YbC$_{6}$ in zero 
magnetic field. Inset: the detailed resistance change near $T_{c}$ (= 6.47 K).}
\label{fig3}
\end{figure}

\newpage

\begin{figure}[h]
\begin{center}
\includegraphics[height=80mm]{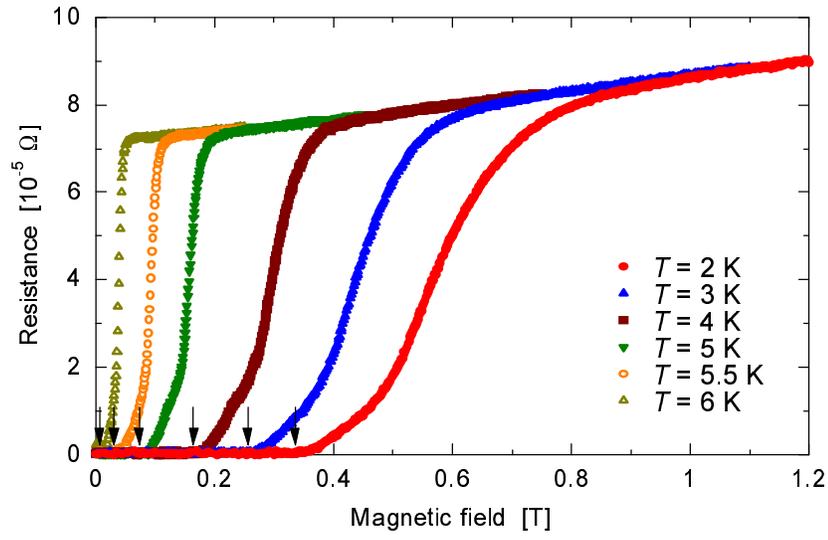}
\end{center}
\caption{The field dependence of resistance in YbC$_{6}$. The field 
direction is parallel to the $ab$-plane. The arrows indicate 
$B_{c2\parallel}$ values where resistance becomes finite with 
increasing field.}
\label{fig4}
\end{figure}

\newpage

\begin{figure}[h]
\begin{center}
\includegraphics[height=110mm]{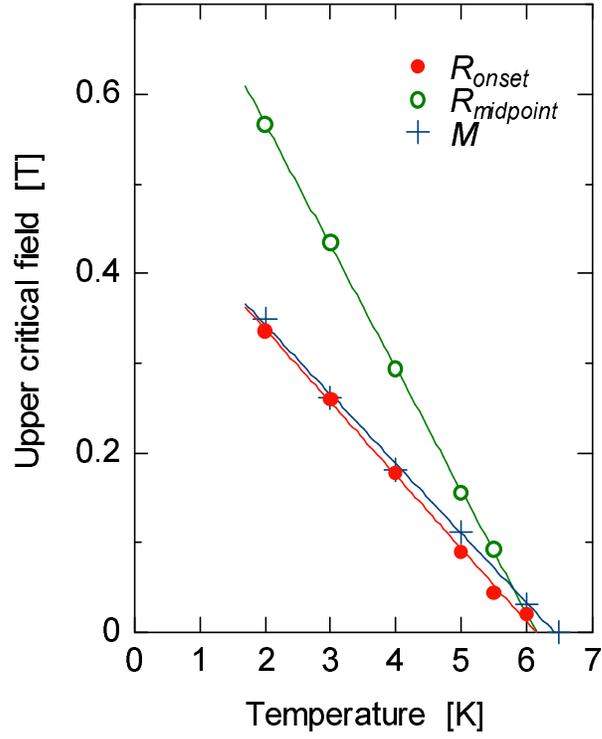}
\end{center}
\caption{The $B_{c2\parallel}$ values of YbC$_{6}$ determined 
from the transport measurements (circles) and the magnetization 
measurements (crosses). The open circles are fields 
defined as midpoints of the resistive transition, which are 
substantially larger than the $B_{c2\parallel}$ values (see text).}
\label{fig5}
\end{figure}

\newpage

\begin{figure}[h]
\begin{center}
\includegraphics[height=85mm]{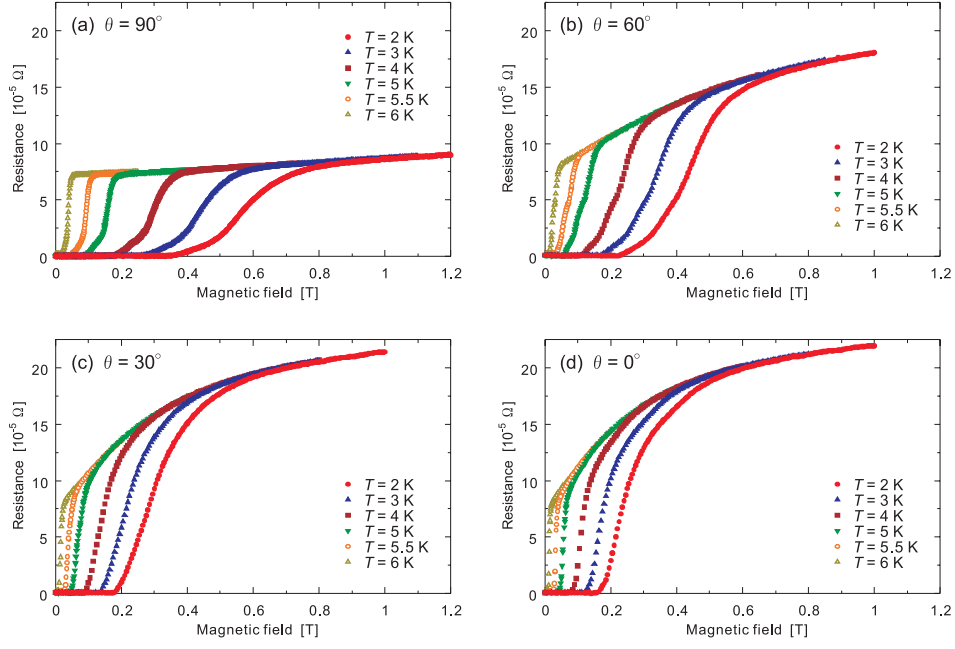}
\end{center}
\caption{The magnetic field dependences of resistance for YbC$_{6}$ 
in various field directions. $\theta$ is defined as an angle 
between the magnetic field direction and the $c$-axis of YbC$_{6}$. 
(a) $\theta$ = 90$^\circ$; (b) $\theta$ = 60$^\circ$; 
(c) $\theta$ = 30$^\circ$; (d) $\theta$ = 0$^\circ$.}
\label{fig6}
\end{figure}

\newpage

\begin{figure}[h]
\begin{center}
\includegraphics[height=110mm]{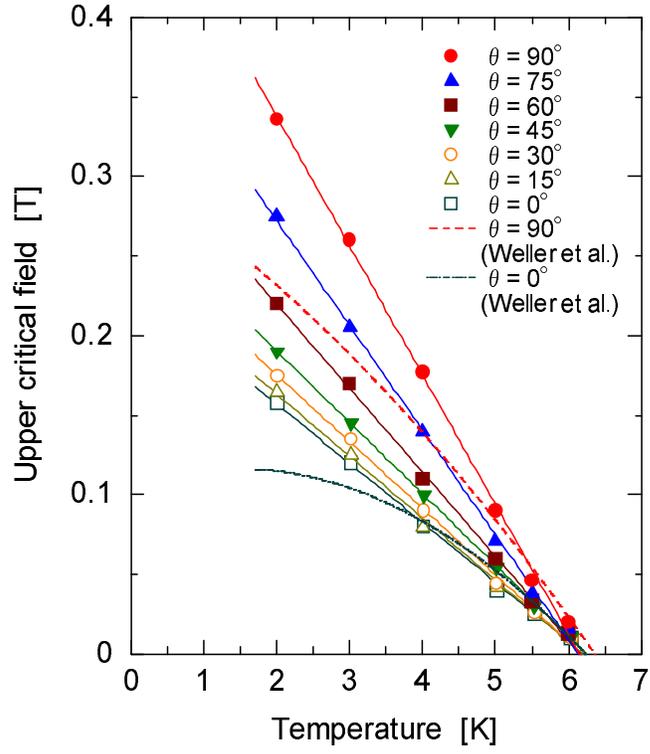}
\end{center}
\caption{The temperature dependences of $B_{c2}$ for YbC$_{6}$ 
in various field directions. The $T$-linear dependence of 
$B_{c2}$ holds at any $\theta$. The dashed lines are 
from Weller et al. \cite{r2}.}
\label{fig7}
\end{figure}

\newpage

\begin{figure}[h]
\begin{center}
\includegraphics[height=90mm]{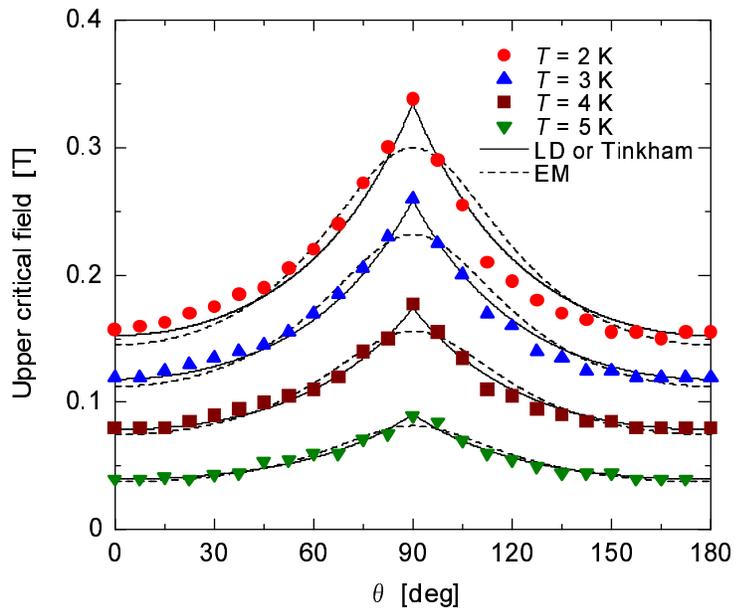}
\end{center}
\caption{The angular dependence of $B_{c2}$ in YbC$_{6}$ 
at various fixed temperatures. Fittings with the EM model 
and the LD or Tinkham model are shown by the dashed lines 
and solid lines, respectively.}
\label{fig8}
\end{figure}

\end{document}